\documentclass[journal]{IEEEtran}

\usepackage{booktabs}
\usepackage{threeparttable}
\usepackage{amsmath,graphicx,cite,amssymb}
\usepackage{amsfonts,multirow,bm,array,setspace,stfloats}

\usepackage{graphicx,float,cite,amssymb}
\usepackage{graphics}
\DeclareGraphicsExtensions{.pdf,.jpeg,.png,.jpg}   
\usepackage{multirow,bm,bbm,array,setspace}
\usepackage{textcomp}
\usepackage{psfrag}
\usepackage{color}
\usepackage{pstricks,enumerate}
\usepackage{bbm}
\usepackage{subfigure}
\usepackage{xpatch}

\usepackage{algorithm,algpseudocode}
\makeatletter
\newcommand{\distas}[1]{\mathbin{\overset{#1}{\kern\z@\sim}}}%
\newsavebox{\mybox}\newsavebox{\mysim}
\newcommand{\distras}[1]{%
	\savebox{\mybox}{\hbox{\kern1pt$\scriptstyle#1$\kern1pt}}%
	\savebox{\mysim}{\hbox{$\sim$}}%
	\mathbin{\overset{#1}{\kern\z@\resizebox{\wd\mybox}{\ht\mysim}{$\sim$}}}%
}

\ExplSyntaxOn
\cs_new:Npn \bibColoredItems #1#2
  {
    \clist_map_inline:nn {#2} { \cs_new:cpn {bib@colored@##1} {#1} } 
  }
\ExplSyntaxOff

\newcommand\bib@setcolor[1]{%
  \ifcsname bib@colored@#1\endcsname
    \expandafter\color\expandafter{\csname bib@colored@#1\endcsname}
  \else
    \normalcolor
  \fi
}

\xpatchcmd\@bibitem
  {\item}
  {\bib@setcolor{#1}\item}
  {}{\PatchFailed}

\xpatchcmd\@lbibitem
  {\item}
  {\bib@setcolor{#2}\item}
  {}{\PatchFailed}

\makeatother
\ifCLASSINFOpdf
\else
\fi
\newtheorem{lemma}{Lemma}
\newtheorem{theorem}{Theorem}

\newcommand{\bd}{\bm d}

\newcommand{\bg}{\bm g}

\newcommand{\bC}{\bm C}

\newcommand{\bx}{\bm{x}}

\newcommand{\by}{\bm{y}}

\newcommand{\bF}{\bm{F}}

\newcommand{\bW}{\bm{W}}

\newcommand{\bz}{\bm{z}}

\newcommand{\bw}{\bm{w}}

\newcommand{\hh}{\mathrm{H}}

\allowdisplaybreaks[4]

\begin{document}
	%\title{Full-Duplex Cellular System with Cross-Channel Transmission and Interference Cancellation}
\title{\huge An Efficient Convex-Hull Relaxation Based Algorithm for Multi-User Discrete Passive Beamforming}
\author{
\IEEEauthorblockN{
    Wenhai Lai, \IEEEmembership{Student Member,~IEEE}, Zheyu Wu, \IEEEmembership{Student Member,~IEEE}, Yi Feng, \IEEEmembership{Student Member,~IEEE}, Kaiming Shen, \IEEEmembership{Senior Member,~IEEE}, and Ya-Feng Liu, \IEEEmembership{Senior Member,~IEEE}
} % <-this % stops a space
%\thanks{
   % Manuscript submitted on \today.

    %Wenhai Lai, Yi Feng, and Kaiming Shen are with the School of Science and Engineering, The Chinese University of Hong Kong (Shenzhen), Shenzhen 518172, China (e-mail: wenhailai@link.cuhk.edu.cn; yifeng1@link.cuhk.edu.cn; shenkaiming@cuhk.edu.cn).

    %Zheyu Wu and Ya-Feng Liu are with the State Key Laboratory of Scientific and Engineering Computing, Institute of Computational Mathematics and Scientific/Engineering Computing, Academy of Mathematics and Systems Science, Chinese Academy of Sciences, Beijing 100190, China (e-mail: \{wuzy, yafliu\}@lsec.cc.ac.cn).
%}
}

		%Telephone: (800) 555--1212, Fax: (888) 555--1212}


\maketitle

\begin{abstract}
Intelligent reflecting surface (IRS) is an emerging technology to enhance spatial multiplexing in wireless networks. This letter considers the discrete passive beamforming design for IRS in order to maximize the minimum signal-to-interference-plus-noise ratio (SINR) among multiple users in an IRS-assisted downlink network. The main design difficulty lies in the discrete phase-shift constraint. Differing from most existing works, this letter advocates a convex-hull relaxation of the discrete constraints which leads to a continuous reformulated problem equivalent to the original discrete problem. This letter further proposes an efficient alternating projection/proximal gradient descent and ascent algorithm for solving the reformulated problem. Simulation results show that the proposed algorithm outperforms the state-of-the-art methods significantly.
\end{abstract}


\section{Introduction}
\IEEEPARstart{I}{ntelligent} reflecting surface (IRS) is an emerging wireless technology that aims to boost signal reception by coordinating the phase shifts of reflected paths \cite{wu2024intelligent}, namely passive beamforming. In practice, the phase shift choice is typically limited to a prescribed discrete set. While the discrete passive beamforming for a single user can be efficiently solved \cite{ren2022linear}, the multiple-user case (e.g., for industrial automation) is much more challenging and remains an open problem. 

The main difficulty in the discrete passive beamforming design for the multi-user case lies in the discrete constraint. Most previous works \cite{wu2019beamforming,li2020joint,you2020channel} just ignore the discrete constraint at the optimization stage and then round the continuous solution to the discrete set. It is generally hard to justify this heuristic relaxation because the relaxed problem without the discrete constraint is not equivalent to the original problem. In sharp contrast to \cite{wu2019beamforming,li2020joint,you2020channel}, this letter advocates a convex-hull relaxation that guarantees the equivalence to the original discrete problem. Further, this letter proposes an efficient alternating projection/proximal gradient descent and ascent algorithm for solving the relaxed problem. As shown in \cite{pan2022overview}, the branch-and-bound method and the heuristic methods including the genetic algorithm have been considered.

Many previous efforts in the realm of passive beamforming focus on the single-user case. To maximize the spectral efficiency for a multiple-input multiple-output (MIMO) link, the work \cite{bahingayi2022low} coordinates the phase shifts of IRS by the Riemannian conjugate gradient method. Considering the double-IRS system, the work \cite{han2020cooperative} proposes a geometric approach to the signal-to-noise ratio (SNR) maximization problem. The more recent work \cite{mei2020cooperative} extends the SNR problem to the case of multiple IRSs. While all the above works require the full channel state information (CSI), a line of studies \cite{Arun2020RFocus, ren2022configuring, Xu2024Coordinating} advocate blind beamforming without using any channel knowledge. Specifically, \cite{Arun2020RFocus} addresses a special {\footnotesize{ON-OFF}} case of passive beamforming, \cite{ren2022configuring} solves the general $K$-ary passive beamforming problem approximately while \cite{ren2022linear} solves it globally, and \cite{Xu2024Coordinating} deals with multiple IRSs, all for maximizing the received signal power at a single target receiver.

For the multi-user case, some existing works consider the common-message multicast network  \cite{tao2020intelligent,huang2020passive,guo2021multiple,yan2023passive}. To avoid channel estimation, the work \cite{tao2020intelligent} proposes an ad-hoc passive beamforming scheme that randomly configures the IRS during the channel coherence interval. Assuming that the CSI is available and the phase shift of each reflective element (RE) can be chosen arbitrarily, the work \cite{guo2021multiple} proposes an alternating direction method of multipliers based method for the max-min SNR problem. In the presence of the discrete constraint, the work \cite{yan2023passive} suggests a gradient descent-ascent (GDA) approach. In contrast, \cite{zheng2021double,xie2020max,ni2021resource,xie2021user,zargari2020energy} account for the downlink transmission with interference. More specifically, the work \cite{zheng2021double} aims to optimize two IRSs jointly in order to maximize the minimum signal-to-interference-plus-noise ratio (SINR) across multiple users, and \cite{xie2020max} considers a weighted max-min SINR problem. Moreover, \cite{ni2021resource} considers maximizing the sum rate, 
\cite{xie2021user} considers minimizing the total latency, and \cite{zargari2020energy} considers maximizing the system energy efficiency. As for the used algorithms, all of
\cite{zheng2021double,xie2020max,ni2021resource,xie2021user} rely on the semidefinite relaxation method while \cite{zargari2020energy} uses the Riemannian gradient method.

\section{System Model}
\label{sec:sys_model}
Consider an IRS-assisted multi-user downlink network in which the base station (BS) with $M$ antennas serves a total of $U$ single-antenna users. The IRS comprises $N$ REs. We use $u=1,2,\ldots,U$ to index the users, and use $n=1,2,\ldots,N$ to index the REs. Denote by $\bd_u\in\mathbb{C}^{M}$ the straight channel from the BS to user $u$. Denote by $\bm F\in\mathbb{C}^{N\times M}$ the channel from the BS to the IRS, whose $(n,m)$th entry is the channel from the $m$th antenna of the BS to the $n$th RE. Denote by $\bg_u\in\mathbb{C}^{1\times N}$ the channel from the IRS to user $u$, whose $n$th entry is the channel from the $n$th RE to user $u$. Let $\bw_u\in\mathbb C^M$ be the beamforming vector at the BS for user $u$; write $\bW=[\bw_1,\bw_2,\ldots,\bw_U]$ with the power constraint $\|\bW\|^2_F\le P$.

We denote by $\theta_n\in[0,2\pi)$ the phase shift induced by RE $n$ in its corresponding reflected path. 
From a practical standpoint, we further assume that each $\theta_n$ can only take on values from the discrete set $\{0,\frac{2\pi}{K},2\times\frac{2\pi}{K},\ldots,(K-1)\times\frac{2\pi}{K}\}$ given some integer $K\ge2$. Define the variable
\begin{equation}
    \bx = [1, x_1, x_2, \ldots, x_N]^\hh\;\text{where each } x_n=e^{j\theta_n}.
\end{equation}
Because of the discrete constraint on $\theta_n$, each $x_n$ is limited to
\begin{equation}
    \label{discrete_set}
    \mathcal{X}=\left\{e^{\frac{j2k\pi}{K}}\mid k=0,1,\ldots,K-1\right\}.
\end{equation}
With the $(N+1)\times M$ matrix
\begin{equation}
    \bm \Phi_u=\left[\begin{array}{cc}
         \bd_u^\top \\
         \mathrm{diag}(\bg_u)\bF
    \end{array}\right],
\end{equation}
the received signal at user $u$ can be computed as $\bx^\hh \bm \Phi_u \bw_u s_u + \sum^U_{u'=1,u'\neq u}\bx^\hh \bm \Phi_u \bw_{u'} s_{u'}+z_u$,
where $s_u\sim\mathcal{CN}(0, 1)$ is the independent symbol intended for user $u$, and 
$z_u\sim\mathcal{CN}(0, \sigma_u^2)$ is the background noise. The resulting SINR of user $u$ is
\begin{equation}
    \mathrm{SINR}_u=\frac{\bx^\hh \bC_{uu}\bx}{\sum_{u'\neq u}\bx^\hh \bC_{uu'}\bx + \sigma^2_u},
\end{equation}
where 
\begin{equation}
\bm C_{uu'} = \bm\Phi_u \bw_{u'}\bw_{u'}^\hh \bm\Phi_u^\hh.
\end{equation}
To achieve the max-min fairness for $U$ downlink users, we consider the joint active and passive beamforming problem as
\begin{subequations}
    \label{discrete problem}
\begin{align}
    \underset{\bW,\,\bx}{\text{maximize}} &\quad \underset{u}{\text{min}}\left\{\mathrm{SINR}_u\right\}
    \label{discrete problem:obj}\\
    \text {subject to} &\quad x_n\in \mathcal{X},\;n=1,2,\ldots,N, \label{discrete_contraint}\\
    &\quad \|\bW\|^2_F\leq P.
\end{align}
\end{subequations}
We propose optimizing two variables $\bW$ and $\bx$ alternatingly (e.g., as in \cite{xie2020max} and \cite{liu2013max}). In particular, when $\bx$ is fixed, the problem of $\bW$ can be optimally solved as in \cite{wiesel2005linear}.

The rest of the letter focuses on optimizing $\bx$ in problem \eqref{discrete problem} with $\bW$ being fixed. The main difficulty in optimizing $\bx$ lies in the discrete constraint \eqref{discrete_contraint}. To overcome this difficulty, many previous works \cite{wu2019beamforming,li2020joint,you2020channel} propose to simply ignore the discrete constraint by allowing each phase shift $\theta_n$ to take an arbitrary value in $[0,2\pi)$, i.e., $\mathcal X$ is relaxed as
\begin{equation}
    \label{uni-circle}
    \hat{\mathcal X} = \big\{e^{j\theta}\mid\theta\in[0,2\pi)\big\}=\big\{x\in\mathbb{C}\mid|x|=1\big\}.
\end{equation}
Hence, each $x_n$ now lies on the unit circle in the complex plane. After the relaxed problem is solved, the obtained solution is rounded to the discrete set $\mathcal X$ in \eqref{discrete_set}. However, such relaxation cannot guarantee the equivalence between the relaxed problem and the original one, thereby resulting in a potentially large performance loss. %\textcolor{blue}{Some other techniques summarized in \cite{pan2022overview} include branch-and-bound method, negative square penalty method and heuristic methods. For the branch-and-bound method and heuristic methods, the former requires exponential complexity and the latter cannot provide any performance guarantee. We will further discuss the negative square penalty method in the next section.}

%We will mainly discuss how to solve problem \eqref{passive_problem} in the subsequent content. Problem \eqref{passive_problem} is difficult to solve for two reasons. First, each $\bm C_{uu}$ and $\bm C_{uu'}$ are the positive semi-definite matrices, so the max-min objective function is not concave. Second, the discrete constraint \eqref{passive_problem:contraint} poses a major challenge. In the existing literature, a common method is to first address the relaxed problem where each $\theta_n$ can be continuously chosen, and then round the solution to the discrete set. This work aims at a much more effective approach to problem \eqref{passive_problem}.

\section{Convex-Hull Relaxation}

\begin{figure}[t]
    \centering
    \includegraphics[width=4.0cm]{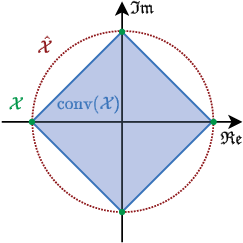}
    \caption{Conventional relaxation $\hat{\mathcal X}$ vs. convex-hull $\mathrm{conv}(\mathcal X)$ when $K=4$.}
    \label{fig:hull_vs_unit}
\end{figure}

Our work also seeks to eliminate the discrete constraint from \eqref{discrete problem}, but by relaxing the discrete set $\mathcal X$ to its convex hull
\begin{equation}
    \label{conv_hull}
    \mathrm{conv}(\mathcal{X})=\left\{\sum_{k=1}^{K}\lambda_k x_k\;\bigg|\; x_k\in\mathcal{X},\sum_{k=1}^{K}\lambda_k=1, \lambda_k\geq 0, \forall\,k\right\}.
\end{equation}
Fig.~\ref{fig:hull_vs_unit} shows the conventional relaxation $\hat{\mathcal X}$ in \eqref{uni-circle} and the convex-hull relaxation $\mathrm{conv}(\mathcal{X})$ in \eqref{conv_hull} of the discrete set $\mathcal X$ when $K=4$. 
The main feature of the proposed convex-hull relaxation $\mathrm{conv}(\mathcal X)$ is that it allows for an equivalent continuous reformulation of \eqref{discrete problem}, which significantly facilitates the algorithmic development. As shown in Fig.~\ref{fig:hull_vs_unit}, all the feasible solutions in $\mathcal X$ turn out to be the vertices of the convex hull $\mathrm{conv}(\mathcal X)$, which are the farthest points from the origin in the complex unit ball. As such, it will encourage choosing a discrete solution from $\mathcal X$ if we require the solution to be sufficiently far away from the origin.

The above idea can be realized by reformulating the discrete problem \eqref{discrete problem} (with respect to $\bx$) as a continuous problem:
\begin{subequations}
\label{penalty_problem}
\begin{align}
    \underset{\bx}{\text{maximize}}&\quad \underset{u}{\text{min}}\left\{\frac{\bx^\hh \bC_{uu}\bx}{\sum_{u'\neq u}\bx^\hh \bC_{uu'}\bx + \sigma^2_u}\right\}+\lambda \|\bx\|_1
    \label{penalty_problem:obj}\\
    \text {subject to} &\quad x_n\in\mathrm{conv}(\mathcal X),\;n=1,2,\ldots,N,
    \label{penalty_problem:contraint}
\end{align}
\end{subequations}
where $\lambda>0$ is a positive parameter. Intuitively, by increasing $\lambda$, the solution of \eqref{penalty_problem} would go toward a vertex of $\mathrm{conv}(\mathcal X)$. Clearly, as $\lambda\rightarrow\infty$, the solution of \eqref{penalty_problem} must be a vertex of $\mathrm{conv}(\mathcal X)$ and hence it must lie in $\mathcal X$. One main result of this letter is to show that it suffices to use a finite parameter $\lambda$ to achieve the above goal, i.e., the relaxed problem \eqref{penalty_problem} with a proper $\lambda$ is equivalent to the original discrete problem \eqref{discrete problem}.

\begin{figure}[t]
\centering
\subfigure[]{
    \label{fig:hull_1}
    \includegraphics[width=2.8cm]{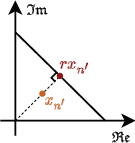}}
\hspace{2em}
\subfigure[]{
    \label{fig:hull_2}
    \includegraphics[width=2.87cm]{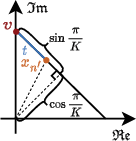}}
\caption{Two key steps in proving Lemma \ref{coro:boundary} and Theorem \ref{prop:global_optimal}. Panel (a) shows the definition of $rx_{n'}$. Panel (b) illustrates the computation of $|x_{n'}|$.}
\label{fig:hull}
\end{figure}

\begin{lemma}
\label{coro:boundary}
All locally and globally optimal solutions of problem \eqref{penalty_problem} lie on the boundary of $\mathrm{conv}(\mathcal X)$, which is denoted as $\partial\mathrm{conv}(\mathcal X)$, so long as the parameter
\begin{equation}
\label{eq:penalty_param_condition}
    \lambda > \max_u \left\{L_u\right\},
\end{equation}
where $L_u = \frac{2\sqrt{N+1}\|\bC_{uu}\|_F}{\sigma_u^2}\left(1+\frac{(N+1)\|\sum_{u^\prime \neq u}\bC_{uu^\prime}\|_F}{\sigma_u^2}\right)$.
\end{lemma}
\begin{IEEEproof}
Suppose that there exists a locally or globally optimal solution $\bx$ of \eqref{penalty_problem} with its entry $x_{n'}\notin\partial\mathrm{conv}(\mathcal X)$. Then, for this $x_{n'}$, we can find a real scalar $r>1$ such that $rx_{n'}\in\partial\mathrm{conv}(\mathcal X)$, as illustrated in Fig.~\ref{fig:hull_1}. For any real scalar $1<\delta<r$, define
\begin{equation}
    \bz^\delta=[1,z_1^\delta,\ldots,z_N^\delta]^\top\;\text{where}\; z_n^\delta=
    \begin{cases}
        \delta x_n,\quad &\text{if }n=n';\vspace{0.2em}\\
        x_n,\quad &\text{otherwise.}
    \end{cases}
\end{equation}
Moreover, consider the function
\begin{equation}
\label{f_u}
    f_u(\bm x) = \frac{\bx^\hh \bC_{uu}\bx}{\sum_{u'\neq u}\bx^\hh \bC_{uu'}\bx + \sigma^2_u}.
\end{equation}
It can be shown that
\begin{align}
\label{chain1}
    f_u(\bx_1)-f_u(\bx_2)
    &\geq -L_u\|\bx_1 - \bx_2\|_2
\end{align}
for any two $\bx_1,\bx_2\in\mathrm{conv}(\mathcal X)$, so $L_u$ is the Lipschitz constant of $f_u(\bx)$. We then have
\begin{equation}
    \label{eq:gap_between_solutions}
    f_u(\bz^\delta) - f_u(\bm x) +\lambda\|\bz^\delta\|_1-\lambda\|\bx\|_1
    \geq \left(\lambda-L_u\right)(\delta-1)|x_{n'}|.
\end{equation}
The relaxed optimization objective in \eqref{penalty_problem:obj} can be written as $f_0(\bx)=\min_u\{f_u(\bx)\}+\lambda\|\bx\|_1$. It immediately follows from \eqref{eq:penalty_param_condition} and \eqref{eq:gap_between_solutions} that $f_0(\bz^\delta) - f_0(\bx)>0$ for all $\delta\in(1,r)$.
This shows that $\bx$ is not a locally optimum solution, so we arrive at a contradiction. The proof is then completed.
\end{IEEEproof}

\begin{theorem}
\label{prop:global_optimal}
The relaxed problem \eqref{penalty_problem} is equivalent to the original discrete problem  \eqref{discrete problem} if the parameter
\begin{equation}
\label{lambda:condition}
    \lambda > \frac{\sin(\pi/K)}{1-\cos(\pi/K)}\cdot\max_u \left\{L_u\right\}.
\end{equation}
\end{theorem}
\begin{IEEEproof}
We start by showing that a solution $\bx$ of problem \eqref{penalty_problem} must be a solution of problem \eqref{discrete problem}. The most nontrivial step is to show that each entry of this solution $\bx$ lies in the discrete set $\mathcal X$. Since any $\lambda$ in \eqref{lambda:condition} must satisfy the condition in \eqref{eq:penalty_param_condition}, $\bx$ must lie on the convex-hull boundary $\partial\mathrm{conv}(\mathcal X)$ according to Lemma \ref{coro:boundary}. Assume that $\bx$ has an entry $x_{n'}\notin\mathcal X$ and $v\in\mathcal X$ is the nearest discrete point, as shown in Fig.~\ref{fig:hull_2}.
% \begin{equation}
%     v = \arg\min_{x\in\mathcal X_K}|x-x_{n'}|.
% \end{equation}
We let
\begin{equation}
    {\bz}=[1,{z_1},\ldots, {z_N}]^\top\;\text{where}\; z_n=
    \begin{cases}
        v, \quad &\text{if }n=n';\vspace{0.2em}\\
        x_n,\quad &\text{otherwise.}
    \end{cases}
\end{equation}
Based on \eqref{chain1}, we further derive
\begin{multline}
    \label{diff between f}
    f_u(\bz) - f_u(\bx) +\lambda\|\bz\|_1-\lambda\|\bx\|_1
    \geq\\ -L_u|v-x_{n'}| + \lambda(1-|x_{n'}|).
\end{multline}
Let $t=|v-x_{n'}|$; observe that $0<t\le\sin(\pi/K)$.
With $K$ and $t$, the value of $|x_{n'}|$ is given by
\begin{equation}
\label{length of x}
    |x_{n'}| = \sqrt{\cos^2\left(\frac{\pi}{K}\right)+\bigg(\sin\left(\frac{\pi}{K}\right)-t\bigg)^2}.
\end{equation}
Substituting \eqref{length of x} into \eqref{diff between f}, we obtain
\begin{align}
    &f_u(\bz) - f_u(\bx) +\lambda\|\bz\|_1-\lambda\|\bx\|_1 \notag\\
    &\geq -L_u \cdot t + \lambda 
   -\lambda\sqrt{\cos^2\left(\frac{\pi}{K}\right)+\bigg(\sin\left(\frac{\pi}{K}\right)-t\bigg)^2}.
   \label{eq:lower_bound}
\end{align}
Observe that when $\bz$ and $\bx$ are both fixed, there exists some $t_0\in(0,\sin(\pi/K)]$ such that the lower bound in the right-hand side of \eqref{eq:lower_bound} increases with $t$ for $t\in(0,t_0)$ and decreases with $t$ for $t\in[t_0,\sin(\pi/K)]$. Furthermore, due to \eqref{lambda:condition}, the infimum of the lower bound must be attained at $t=0$, so it holds that $f_u(\bz) - f_u(\bx) +\lambda\|\bz\|_1-\lambda\|\bx\|_1>0$ for any $t\in(0,\sin(\pi/K)]$.
We still write the objective function of \eqref{penalty_problem} as $f_0(\bx)=\min_u\{f_u(\bx)\}+\lambda\|\bx\|_1$. It follows that $f_0(\bz)>f_0(\bx)$. The above result contradicts the assumption that $\bx$ is globally optimal for problem \eqref{penalty_problem}. Therefore, as long as $\bx$ is a globally optimal solution of \eqref{penalty_problem}, we must have $x_n\in\mathcal X$ for all $n$ and $\|\bx\|_1=N+1$. Now it is simple to show that the solution $\bx$ of \eqref{penalty_problem} must be a solution of \eqref{discrete problem}. For the contradiction purpose, assume that $\bx$ is not optimal for \eqref{discrete problem} while $\tilde\bx$ is a globally optimum solution of \eqref{discrete problem}. Thus, we must have $\min_u \{f_u(\tilde\bx)\}>\min_u \{f_u(\bx)\}$. Moreover, because $\|\tilde\bx\|_1=\|\bx\|_1=N+1$, we further have $f_0(\tilde\bx)>f_0(\bx)$; which contradicts the assumption that $\bx$ is globally optimal for \eqref{penalty_problem}. Thus, $\bx$ must be a globally optimal solution of \eqref{discrete problem}.

We then verify the converse, i.e., a solution $\bx$ of \eqref{discrete problem} must be a solution of \eqref{penalty_problem}. Suppose that $\tilde\bx$ is a solution of \eqref{penalty_problem}. From the previous part, we know that $\tilde\bx$ must be a solution of \eqref{discrete problem}. Since $\bx$ and $\tilde\bx$ are both feasible for \eqref{discrete problem}, it follows that $\|\bx\|_1=\|\tilde\bx\|_1=N+1$; since $\bx$ and $\tilde\bx$ are both globally optimal for \eqref{discrete problem}, it holds that $\underset{u}{\text{min}}\{f_u(\tilde\bx)\}=\underset{u}{\text{min}}\{f_u(\bx)\}$; thus, $\bx$ is equally good as $\tilde\bx$ for \eqref{penalty_problem}, so $\bx$ must be a globally optimal solution of \eqref{penalty_problem}. Combining the above results verifies the equivalence.
\end{IEEEproof}

Here are two remarks on Lemma \ref{coro:boundary} and Theorem \ref{prop:global_optimal}. First, the use of $\|\bx\|_1$ in \eqref{penalty_problem} is critical. If we use $\|\bx\|_2^2$ instead, then it is difficult to show that all the locally optimal points must lie on the convex hull boundary. Further, if we switch to some other penalty terms, then we are faced with the problem of whether a finite lower bound on $\lambda$ still exists. Second, using the convex hull to approximate the discrete set can be found in the literature \cite{wu2023efficient,liu2024survay}, but its application to the passive beamforming has not yet been explored. Moreover, because the objective function in \eqref{discrete problem} is more complicated than that in \cite{wu2023efficient}, it requires new efforts to establish the equivalence to the original problem, as shown in Lemma \ref{coro:boundary} and Theorem \ref{prop:global_optimal}. 

\begin{figure}[t]
    \centering
    \includegraphics[width=7cm]{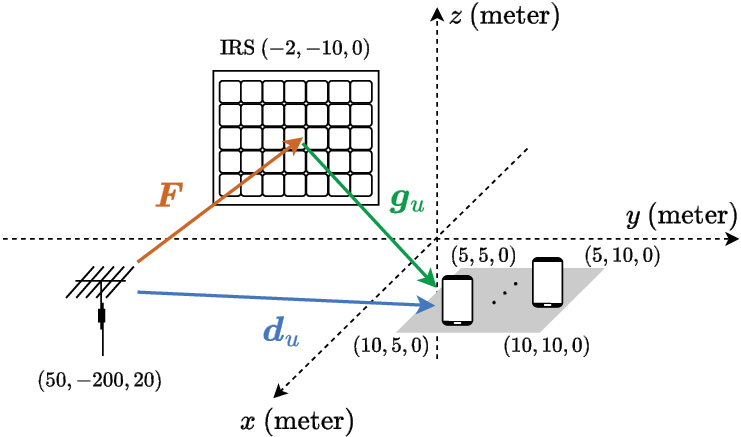}
    \caption{The IRS-assisted downlink network in our simulations.}
    \label{fig:simu_loc}
\end{figure}

\begin{figure*}[t]
\centering
    \begin{minipage}[t]{0.3\linewidth}
        \centering
    \includegraphics[width=6.2cm]{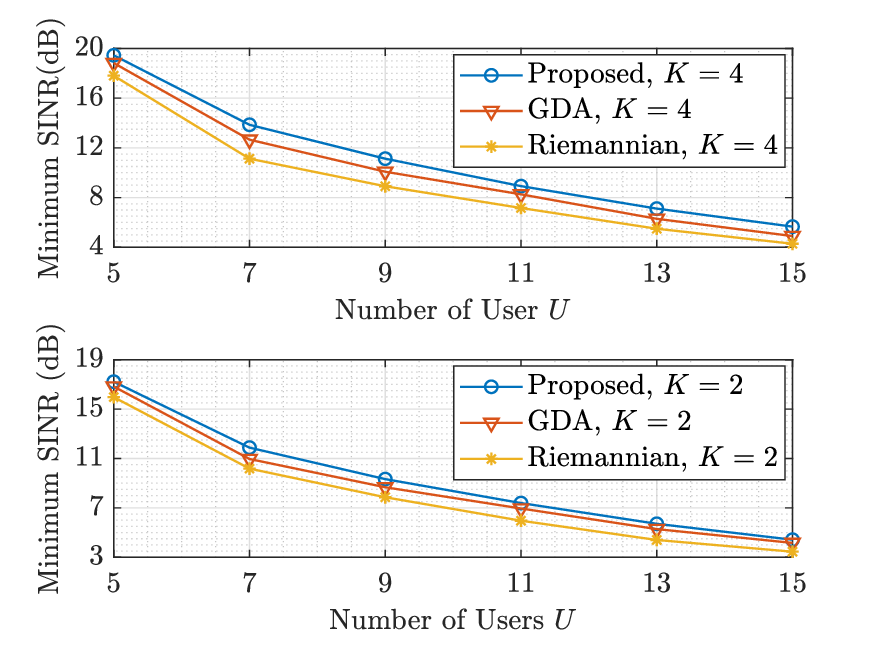}
    \caption{Minimum SINR vs. $U$ when $N=500$.}
    \label{fig:SNR_vs_U}
    \end{minipage}
    \quad
    \begin{minipage}[t]{0.3\linewidth}
        \centering
        \includegraphics[width=6.2cm]{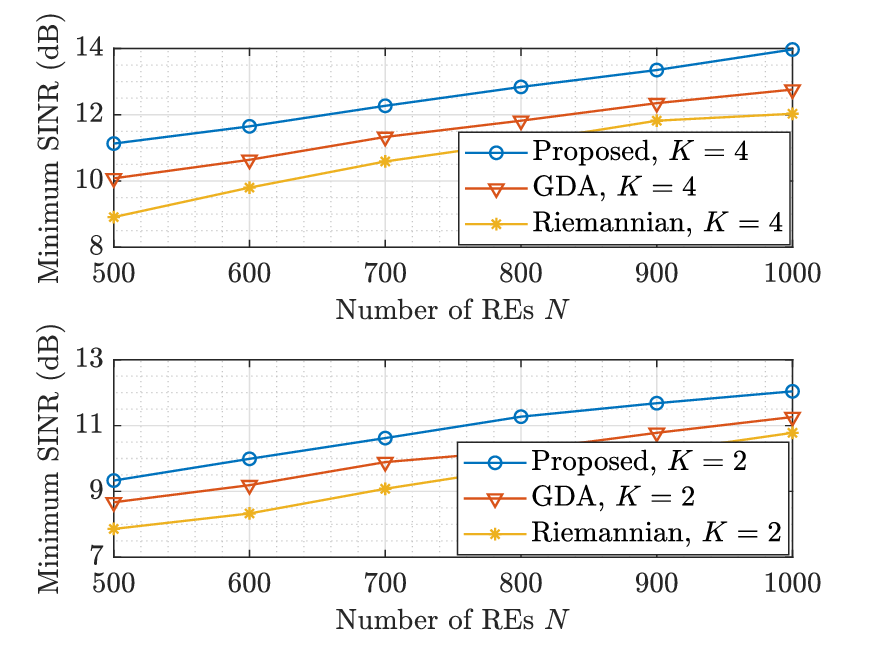}
    \caption{Minimum SINR vs. $N$ when $U=9$.}
    \label{fig:SNR_vs_N}
    \end{minipage} %\hfill
    \quad
    \begin{minipage}[t]{0.3\linewidth}
        \centering
    \includegraphics[width=6.2cm]{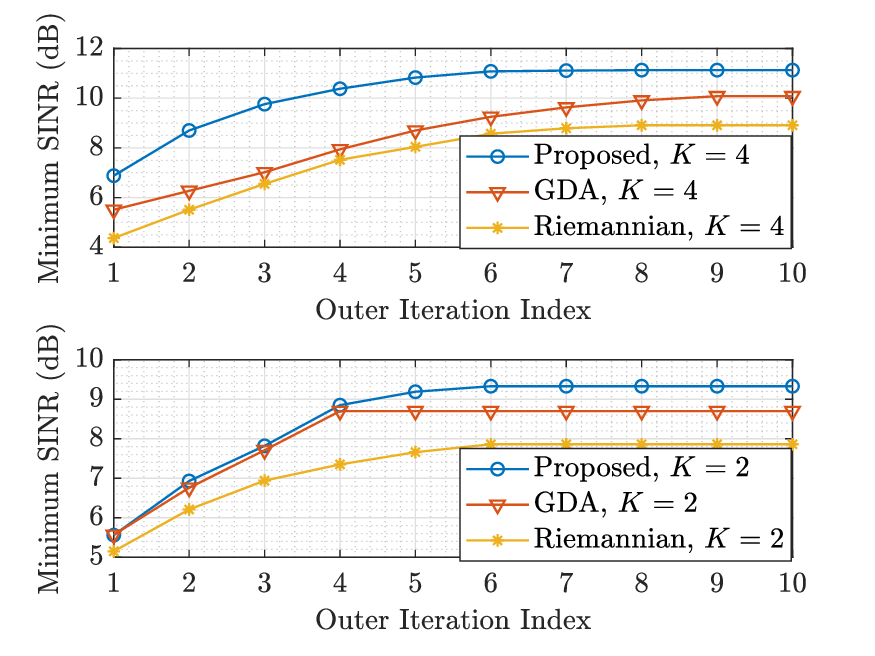}
    \caption{Convergence as $N=500$ and $U=9$.}
    \label{fig:Convergence}
    \end{minipage}
\end{figure*}

\section{Proposed Algorithm}
\label{subsec:APGDA}

\begin{algorithm}[t]
\caption{Proposed Algorithm}
\label{alg:APGDA}
\begin{algorithmic}[1]
    \State\textbf{input:} $\{\bm\Phi_u\}$ and $\mathcal{X}$.
    \For{$i=1,2,\ldots,T_1$}
    \State Decide $\bW$ by the method in \cite{wiesel2005linear}.
    \State initialize $\lambda=0.0001$.
    \For{$j=1,2,\ldots,T_2$}
    \If{mod$(j, 100)$ is 0}
    \State Update $\lambda=10\lambda$.
    \EndIf
    \State Update $\by$ by \eqref{eq:y_update}.
    \State Update $\bx$ by \eqref{eq:x_update}.
    \EndFor
    \EndFor
    \State \textbf{Output:} $\bW$ and $\bx$.
\end{algorithmic}
\end{algorithm} 
% The nonsmoothness of problem \eqref{penalty_problem} rules out the conventional gradient method. A natural idea is to apply the proximal gradient method. As a well-known result in this field, the proximal gradient method can be carried out efficiently when the nonsmooth part of the optimization objective is a 1-norm term. Thus, we wish to further convert the optimization objective in  \eqref{penalty_problem:obj} to a smooth function plus the nonsmooth 1-norm term $\lambda\|\bx\|_1$. Toward this end, we introduce an auxiliary variable $\by\in\mathbb R^U$ and recast problem \eqref{penalty_problem} to

In this section, we propose an alternating projection/proximal gradient descent and ascent algorithm for solving problem \eqref{penalty_problem}. The proposed algorithm is based on the following reformulation of problem \eqref{penalty_problem}:

\begin{subequations}
\label{equivalent_problem}
\begin{align}
    &\max_{\bm x}\min_{\bm y\in\bm\triangle}\quad\sum_{u=1}^U\frac{y_u\bx^\hh \bC_{uu}\bx}{\sum_{u'\neq u}\bx^\hh \bC_{uu'}\bx + \sigma^2_u}+\lambda \|\bx\|_{1}
    \label{equivalent_problem:obj}\\
    &\text {subject to} \quad\, x_n\in\mathrm{conv}(\mathcal X),\;n=1,2,\ldots,N,
\label{equivalent_problem:contraint}
\end{align}
\end{subequations}
where $\by=[y_1,y_2,\ldots,y_U]^\top$ is an auxiliary variable and $\bm\triangle=\{\by\mid\bm 1^\top\bm y = 1,\, y_u \geq 0,\,\forall u\}$ is the simplex. The equivalence between \eqref{penalty_problem} and \eqref{equivalent_problem} follows from the following observation: if $\bx$ has been optimally determined, we would assign $y_u=1$ to the minimum term while setting $y_{u'}=0$ for the rest.

We propose optimizing $\bx$ and $\by$ alternatingly in \eqref{equivalent_problem}. Denote by $g(\bx,\by)$ the first smooth term and denote by $h(\bx)$ the second nonsmooth term of the optimization objective in \eqref{equivalent_problem:obj}, respectively. For fixed $\bx$, the auxiliary variable $\by$ can be updated by performing a projection gradient descent step:
\begin{align}
\label{eq:y_update}
\bm y^{t+1} = \mathrm{Proj}_{\bm \triangle} \Big(\bm y^t - \alpha^t \nabla_{\bm y}g(\bm x^t, \bm y^t)\Big),
\end{align}
where $t$ is the iteration index, $\alpha^t>0$ is the stepsize, and $\mathrm{Proj}_{\bm \triangle}(\cdot)$ is the projection onto $\bm\triangle$. The simplex projection can be efficiently done with a complexity of $O(U\log U)$ \cite{condat2016fast}.
For fixed $\by$, update $\bx$ by the proximal gradient ascent:
\begin{equation}
    \bm x^{t+1}=\arg\min_{\bm x\in\mathcal A}\frac{1}{2\beta^t}\lVert\bm x-\bm x^t-\beta^t\nabla_{\bm x} g(\bm x^{t}, \bm y^{t+1})\rVert_2^2 - h(\bm x),\label{eq:x_update}
\end{equation}
where $\beta^{t}>0$ and $\mathcal A=\big\{\bx:\text{each } x_n\in\mathrm{conv}(\mathcal X)\big\}$. The operation in \eqref{eq:x_update} can be done efficiently as in \cite{wu2023QCE}. Algorithm \ref{alg:APGDA} summarizes the above steps. The per-iteration complexity of our algorithm is $\mathcal{O}(N^2U^2)$, and that of GDA  \cite{yan2023passive} is $\mathcal{O}(N^2U^2+NK)$, where $N\gg U$. Note that the complexity is at least quadratic in $N$ if the gradient is used.

The above algorithm can be extended to the weighted sum-rate maximization problem. Our algorithm can also be extended to the active IRS case \cite{zhi2022active}. In this case, we perform the proposed algorithm and update the amplification factor as in \cite{zhi2022active} in an alternating fashion.
% The lower bound on $\lambda$ then needs to be reconsidered.

\section{Simulation Results}
Fig.~\ref{fig:simu_loc} shows the considered IRS-assisted multi-user downlink network at $2.6$ GHz. The transmit antennas are arranged as a uniform linear array while the REs are arranged as a uniform planar array, with half-wavelength spacing. The pathloss model follows \cite{jiang2021learning}. We adopt the 
Rician fading model \cite{david2005fundamentals} with the Rician factor $k=10$. %\textcolor{blue}{In our simulation, all results are generated by avaraging out 1000 realizations of fading channels.}

Let $P=30$ dBm, $\sigma^2_u=-90$ dBm, $K\in\{2,4\}$, and $M=64$. The users are randomly distributed in the shaded area as shown in Fig.~\ref{fig:simu_loc}. The GDA method in \cite{yan2023passive} and the Riemannian method in \cite{zargari2020energy} are the benchmarks; the solution of the Riemannian method is rounded to the discrete set. Note that our algorithm contains the outer iteration that updates $\bW$ iteratively as well as the inner loop that updates $\bx$ and $\by$ iteratively under the current $\bW$. We run 1000 inner iterations per outer iteration. At the beginning of each outer iteration, $\lambda$ is initialized to $0.0001$, and then is multiplied by $10$ every after $100$ inner iterations. Moreover, initialize both $\alpha$ and $\beta$ to $0.01$, and update them as $\alpha^{t+1}=0.997\alpha^t$ and $\beta^{t+1}=0.997\beta^t$.%; this is also the stepsize setting for GDA and Riemannian algorithms.

\begin{table}[t]
\small
\renewcommand{\arraystretch}{1.3}
\centering
\caption{\small Running Time of Different Methods When $N=500$.}
\begin{tabular}{lrrrr}
\firsthline
& \multicolumn{3}{c}{Running Time (second)}\\
\cline{2-4}
Method      & $U=5$ & $U=10$ & $U=15$\\
\hline
Proposed Algorithm   & 91.1 & 115.3 & 151 \\
GDA \cite{yan2023passive}    & 98.8 & 171.3 & 221.2 \\
Riemannian \cite{zargari2020energy} & 96.9 & 153.4 & 194.5 \\
\lasthline
\end{tabular}
\label{tab:run_time}
\end{table}

Fig.~\ref{fig:SNR_vs_U} shows the minimum SINR versus the number of users $U$ when $N=500$. Observe that the minimum SINRs achieved by the different algorithms all decrease with $U$; this implies that it is increasingly difficult to coordinate the beams when more users are in the network. Observe also that our algorithm always outperforms the GDA and Riemannian algorithms.
Fig.~\ref{fig:SNR_vs_N} shows the minimum SINR versus the number of REs $N$ when $U=9$. As one can expect, the minimum SINRs achieved by all the algorithms increase with $N$. Moreover, we compare the proposed algorithm and the exhaustive search method for a toy model with $N=10$ and $U=5$. Our simulation shows that the proposed algorithm attains the minimum SINR of 7.31 dB while the global optimum achieved by the exhaustive search equals 7.41 dB.

Moreover, we compare the convergence behaviors of the different algorithms in Fig.~\ref{fig:Convergence}. Observe that our algorithm has a much faster convergence. Note that the convergence rate of GDA is sensitive to $K$. Finally, observe from Table \ref{tab:run_time} that our algorithm has the highest time-efficiency.

\section{Conclusion}
\label{sec:Conclusion}
This letter aims to maximize the minimum SINR among multiple users in an IRS-assisted downlink network under a \emph{discrete} constraint on passive beamforming. While many existing works adopt a unit-circle relaxation, this letter proposes a convex-hull relaxation so that the new continuous problem is guaranteed to be equivalent to the original discrete problem. The relaxed problem can be efficiently solved by the alternating projection/proximal gradient descent and ascent algorithm.

% \bibColoredItems{blue}{pan2022overview, zhi2022active}
\bibliographystyle{IEEEtran}     
\bibliography{IEEEabrv,Ref}
\end{document}